\documentstyle[preprint,prd,aps]{revtex}
\begin{document}
\draft

\title {O(d, d)--Symmetry and Ernst Formulation for\\
Einstein--Kalb--Ramond Theory in Three Dimensions}

\author{Alfredo Herrera}

\address{Joint Institute for Nuclear Research,\\
Dubna, Moscow Region 141980, RUSSIA, \\
e-mail: alfa@cv.jinr.dubna.su}

\author {\rm and}

\author{Oleg Kechkin}

\address{Nuclear Physics Institute,\\
Moscow State University, \\
Moscow 119899, RUSSIA, \\
e-mail: kechkin@cdfe.npi.msu.su}

\date{December 1996}

\maketitle

\draft
\begin{abstract}
The Ernst--like matrix representation of the multidimensional
Einstein--Kalb--Ramond theory is developed and the O(d,d)--symmetry
is presented in the matrix--valued $SL(2,R)$--form. The analogy with
the Einstein and Einstein--Maxwell--Dilaton--Axion
theories is discussed.
\end{abstract}

\draft

\narrowtext
\section{Introduction}
Recently much attention had been attracted to the study of symmetries of the
dimensionally reduced low energy effective string theory.
Such a theory describes the gravitational, dilaton, Kalb--Ramond and a
set of $n$ Abelian vector fields in $D + d$ dimensions. It becomes
$O(d, d+n)$--invariant after the compactification of $d$ dimensions on a
torus  \cite {ms} and allows the $O(d, d+n)/O(d) \times O(d+n)$ coset
matrix formulation in the case of $D = 3$ \cite {s1}.
The transformations of the $O(d, d+n)$ group were explored for the generation of
new solutions in this theory as well as for the prediction of the features
of exact string excitations \cite {s2}--\cite {tc}.

The four--dimensional Einstein--Maxwell--Dilaton--Axion (EMDA) theory,
being the simplest model of this type, admits instead of the orthogonal
representation the
symplectic one \cite {gk1}.
It was established that this system
possesses the $Sp(4,R)$ symmetry group and allows the
coset $Sp(4, R)/U(2)$ matrix representation in the stationary case. In
\cite {gk2} the K\"ahler formulation for this model,
which generalizes the well known Ernst formalism for the stationary Einstein
system, was established. Finally, in \cite {ky1} it was established the complete
formal analogy between EMDA and vacuum Einstein theories in the stationary
and stationary axisymmetric cases.

In this paper we show that the $O(d, d)$ symmetric ($n=0$) effective heterotic
string theory, which describes the Einstein and Kalb--Ramond (EKR) fields,
admits classical procedures well known for the stationary Einstein
system. Namely, the EKR relations directly map to the Einstein ones under
the change of the symmetric matrix $G$, constructed on the additional
Kaluza--Klein
metric components, by the real part of the vacuum Ernst potential $f$, and
the antisymmetric Kalb--Ramond matrix $B$ by its imaginary part
$i\chi$, correspondingly.
We construct the matrix scalar--vector Lagrange representation of
the model which relates with the original metric (non--target space)
formulation of the Einstein theory.

Following Maharana and Schwarz we unite the metric and Kalb--Ramond matrices
into the $d \times d$ matrix $X = G + B$ which provides the Ernst--like
formulation of the problem. It is shown that in terms of this matrix the target
space duality group becomes the matrix--valued $SL(2, R)$ one. Its three
subgroups are identified as the matrix generalizations of the gauge shift,
rescaling and Ehlers transformations for the vacuum Einstein theory. The
consequences of the imposition of the additional axisymmetric property will be
presented in the following publications.
\section{Chiral Matrix Formulation}
In this paper we discuss the system with the action
\begin{equation}
{\cal S} = \int d^{3+d}x {\mid {\cal G} \mid}^{\frac {1}{2}} \left\{ - {\cal R} +
\frac {1}{12} {\cal H}^2 \right\},
\end{equation}
where ${\cal R}$ is the Ricci scalar for the metric ${\cal G}_{M N}$,
$(M = 0, ..., 3 + d)$
and
\begin{equation}
{\cal H}_{MNL} = \partial _{M} {\cal B}_{NL}
+ {\rm cyc.\,\, perms.}
\end{equation}
Such a system arises in the low energy limit of heterotic string theory or in
the frames of $N=1$ supergravity. It does not contain the dilaton and gauge
vector fields, so that this theory coincides with the $(3 + d)$--dimensional
Einstein--Kalb--Ramond one.

Following Maharana and Schwarz \cite {ms}, one can extract $d$
dimensions using the Kaluza--Klein compactification on a torus. Doing so
one obtains the
$O(d, d)$--symmetric $\sigma$--model action describing 3--dimensional
gravity coupled with a set of scalar and Abelian vector fields
\cite {s1}.
(The Sen's formulae can be transformed into the our ones using the replacement
$7 \rightarrow d$ and $16 \rightarrow n$.)
In this work we study the case when the vector fields are not present,
i.e., we suggest that the metric and Kalb--Ramond field components with mixed
indeces are equal to zero:
\begin{equation}
{\cal G}_{\mu, n + 2} = {\cal B}_{\mu, n + 2} = 0,
\end{equation}
$\mu = 0, ..., 2; \, n = 1, ..., d$.
It is easy to see that such a restriction does not provide any constraints
on the remainder variables and can be considered as a non--trivial ansatz
for the EKR theory.

In \cite {ms} and \cite {s1} it is shown that the result of the Kaluza--Klein
compactification can be represented using the following 3--dimensional
effective action:
\begin{eqnarray}
^3 S =
\int d^3 x {\mid g \mid}^{\frac {1}{2}} \left\{ - ^3 R +
\frac {1}{8} Tr\left[(J^M)^2\right]
\right\},
\end{eqnarray}
where $J^M = \nabla M \, M^{-1}$.
Here the curvature scalar $^3 R$ is constructed on the 3--metric
$g_{\mu \nu} = {\cal G}_{\mu \nu}$ and the chiral matrix $M$ is defined
as
\begin{eqnarray}
M = \left (\begin{array}{crc}
G^{-1} &\quad & - G^{-1}B\\
BG^{-1} &\quad & G - BG^{-1}B\\
\end{array}\right ),
\end{eqnarray}
with $G_{mn} = {\cal G}_{m + 2, n + 2}$ and $B_{mn} =
{\cal B}_{m + 2, n + 2}$ (thus $G^T = G, \, B^T = - B$). This matrix
satisfies the relations
\begin{eqnarray}
M^T = M, \qquad M \eta M = \eta, \qquad {\rm where} \qquad
\eta = \left (\begin{array}{crc}
0 & I_d\\
I_d & 0\\
\end{array}\right ),
\end{eqnarray}
i. e., belongs to the coset $O(d, d)/O(d) \times O(d)$ \cite {ms}.

It is easy to see that the form of the Gauss decomposition (5) is
very similar to the ones of Einstein and
Einstein--Maxwell--Dilaton--Axion theories \cite {ky1}. Moreover,
the case of an arbitrary value of $d$ corresponds to the general case of the
$Sp(2d, R)/U(d)$ chiral matrix \cite {ky2}. The only difference is connected with
the sign ``-'' in the upper right part of (5) and with the antisymmetric
character of the matrix $B$. As it will be shown below, the analogy between EKR
theory and symplectic systems permits to perform all the classical
procedures well known for the stationary Einstein theory.

The action (4) allows the symmetry transformations belonging to the
group $O(d, d)$.
Namely, any matrix $C \in O(d, d)$, i. e. satisfying the relation
$C^T \eta C = \eta$, defines an automorphism
\begin{equation}
M \rightarrow C^T M C
\end{equation}
for the target space of the problem. The explicit form of a group matrix
which can be continuously transformed to the unit one is:
\begin{eqnarray}
C = \left (\begin{array}{crc}
(S^T)^{-1} & \quad & - (S^T)^{-1} R\\
- L (S^T)^{-1} & \quad & S + L(S^T)^{-1}R\\
\end{array}\right ),
\end{eqnarray}
where $R^T = - R$ and $L^T = - L$. As an example of $O(d, d)$ transformation
which can not be parametrized according to (8), one can take the transformation
defined by the matrix $\eta$. It corresponds to the recently discussed
strong--weak coupling duality transformation $M \rightarrow M^{-1}$
(\cite {s1} and \cite {s3}) which formally exists for any chiral matrix.

It must be noted that the EKR theory with
${\cal G}_{\mu, 2 + m} \neq 0$ and ${\cal B}_{\mu, 2 + m} \neq 0$ is
also $O(d, d)$--symmetric \cite {ms}, but only the ansatz under consideration
allows the formalism developed below.
\section{Ernst Matrix Potential}
The set of Euler--Lagrange equations corresponding to the 3--action (4)
\begin{equation}
\nabla J^M = 0,
\end{equation}
\begin{equation}
^3R_{\mu \nu} = \frac {1}{8} Tr\left[J^M_{\mu}J^M_{\nu}\right],
\end{equation}
can also be written in terms of the matrices $G$ and $B$:
\begin{equation}
\nabla J^B - J^G J^B = 0,
\end{equation}
\begin{equation}
\nabla J^G - (J^B)^2 = 0,
\end{equation}
\begin{equation}
^3R_{mn} = \frac {1}{4}Tr\left[J^G_{\mu}J^G_{\nu} - J^B_{\mu}J^B_{\nu}\right],
\end{equation}
where $J^G = \nabla G \, G^{-1}$ and $J^B = \nabla B \, G^{-1}$.
One can prove that these equations correspond to the action
\begin{eqnarray}
^3 S =
\int d^3x {\mid g \mid}^{\frac {1}{2}} \left\{ - ^3 R +
\frac {1}{4} Tr\left[(J^G)^2 - (J^B)^2\right]
\right\}.
\end{eqnarray}

This action, as well as Eqs. (11)--(13), can be directly transformed into the
stationary Einstein ones by the replacement
\begin{equation}
G \rightarrow f, \qquad B \rightarrow i\chi,
\end{equation}
where $f$ is the $g_{tt}$ metric coefficient and $\chi$ is the rotational
potential. It is  well known that the Ernst potential
$E = f + i\chi$ \cite {e}
plays an important role in the Einstein theory. It
provides the K\"ahler formulation for the stationary Einstein system \cite {m},
and allows to represent the stationary axisymmetric field configurations in
compact form. Using the substitution (15) one can introduce the
analogy of the Ernst potential of the Einstein theory for the EKR one:
\begin{equation}
X = G + B,
\end{equation}
which firstly was considered in \cite {ms}.

It can easily be verified that Eqs. (11)--(13) can be rewritten as
\begin{equation}
\nabla J^X = J^X\left(J^X + J^{X^T}\right),
\end{equation}
\begin{equation}
^3 R_{\mu \nu} = Tr\left(J^X_{(\mu}J^{X^T}_{\nu)}\right),
\end{equation}
where $J^X = \nabla X (X + X^T)^{-1}$.
This system is very similar to the Ernst one \cite {e}--\cite {ky2}
for the symplectic theories and so, the matrix function $X$ can be called
``matrix Ernst potential'' by natural reasons.

The corresponding to Eqs. (17) and (18) action has the form
\begin{equation}
^3 S = \int d^3x {\mid g \mid}^{\frac {1}{2}}\left\{ - ^3 R +
Tr(J^X J^{X^T})\right\}.
\end{equation}
One can see that the complex conjugation in the Einstein case transforms
into the transposition in the EKR one. The symmetries of this action can be
directly obtained from the formulae (5), (7) and (8) for the chiral matrix $M$.
As result one obtains that the transformation
\begin{equation}
X \rightarrow S^T[X^{-1} + L]^{-1}S + R
\end{equation}
has the same sense that Eq. (7). Thus, the complete set of isometry
transformations of EKR theory has a matrix--valued $SL(2, R)$ form. This
allows to understand the subgroups determined by matrices $R$, $S$ and
$L$ as the matrix analogies of gauge shift, rescaling and Ehlers
transformations, correspondingly \cite {k}--\cite {eh}.
\section{Dualization Procedure}
Now let us establish an alternative Lagrange formulation of the problem
based on the use of non--target space variables. One can see that
Eq. (11), being rewritten as
\begin{equation}
\nabla [G^{-1}(\nabla B)G^{-1}] = 0,
\end{equation}
ensures the compatibility conditions for the relation
\begin{equation}
\nabla \times \vec \Omega = G^{-1}(\nabla B)G^{-1}
\end{equation}
which defines the antisymmetric vector matrix $\vec \Omega$.
This matrix provides another representation of the EKR system corresponding
to the original metric one of the case of Einstein theory
\cite {iw} and \cite {k}, i. e., the
formulation with the metric components $\omega _i = {g_{tt}}^{-1}g_{ti}$.

It is easy to see from Eq. (22) that the matrices
$G$ and $\vec \Omega$ satisfy the relation
\mbox {$\nabla \times [G \,\,(\nabla \times \vec \Omega) \,\,G] = 0$}
which also can be rewritten using the matrix current $J^{\vec \Omega} =
G \,\,\nabla \times \vec \Omega$ as
\begin{equation}
\nabla \times J^{\vec \Omega} - J^{\vec \Omega}\times J^G = 0.
\end{equation}
This relation together with Eqs. (12) and (13), expressed in terms
of the matrices $G$ and $\vec \Omega$
\begin{equation}
\nabla J^G - \left(J^{\vec \Omega}\right)^2 = 0,
\end{equation}
\begin{equation}
^3 R_{\mu \nu} =\frac {1}{4} Tr\left(J^G_{\mu} J^G_{\nu} -
J^{\vec \Omega}_{\mu}J^{\vec \Omega}_{\nu}\right)
\end{equation}
form the complete set of the Euler--Lagrange equations for the action
\begin{equation}
^3 S = \int d^3x {\mid h \mid}^{\frac {1}{2}}\left\{- ^3 R + \frac {1}{4}
Tr\left[(J^G)^2 + (J^{\vec \Omega})^2\right]\right\};
\end{equation}
thus these matrices provide an alternative Lagrange formulation of the
theory under consideration. A similar procedure for the symplectic
$Sp(2d, R)/U(d)$ theories has been performed in \cite {ky2}.

At the end of the paper we would like to remark that the low
energy limit of heterotic string theory with dilaton and $n$ Abelian vector
fields, being reduced to three dimensions \cite {s1}, also allows a dualization
procedure. As it can be shown, this procedure is more similar to the
Einstein--Maxwell one, than to discussed here Einstein--like construction.
The reason of such a remarkable circumstance is related with the closed
analogy between the \mbox{$O(d, d+n)/O(d) \times O(d+n)$}
chiral matrix of the string system  with the
\mbox{$SU(1, 2)/U(1) \times SU(2)$}
coset matrix of Eistein--Maxwell theory \cite {gek}.
\section{Conclusion}
It is easy to see that Einstein--Kalb--Ramond system allows all the classical
procedures established for the Einstein and Einstein--Maxwell--Dilaton--Axion
theories in the stationary axisymmetric case. Namely, in the forthcoming
publications it will be shown that this system, being reduced to two dimensions,
also allows the Kramer--Neugebauer transformation
(\cite {kn}, \cite {ky1} and \cite {ky2}), the alternative
(Belinsky--Zakharov--like \cite {bz}) chiral matrix formulation and, moreover,
the construction of the Geroch group \cite {g1}--\cite {g2}
and the O(d, d) matrix analogy of the
Hauser--Ernst formalism \cite {he1}--\cite {wg}.
\acknowledgements
We would like to thank our colleagues from the JINR and NPI
for an encouraging relation to our work. One of the authors (A. H.) would like
to thank CONACYT and SEP for partial financial support.

\end{document}